\documentstyle[aps,prd,epsfig]{revtex}
\def\DD{{\mathrm d}}

\def\Journal#1#2#3#4{{#1}{\bf #2}, #3 (#4)}

\def\NPB{{Nucl. Phys.} \bf B}
\def\PLB{{Phys. Lett.} \bf B}

\def\PRD{{Phys. Rev.} \bf D}

\def\ZPC{{Z.~Phys.} \bf C}
\begin{document}
\draft
\title{$\Delta G(x,\mu^2)$ from jet and prompt photon production at {\sc Rhic}}
\author{O.~Martin and A.~Sch\"afer}
\address{Institut f\"ur Theoretische Physik, Universit\"at Regensburg,
D-93040~Regensburg, Germany}
\date{\today}
\maketitle
%

\begin{abstract}
We compare theoretical results for jet and prompt photon production
in $\vec p\vec p$-collisions at {\sc Rhic} obained with three different methods: a) the 
unpolarized event generator {\sc Pythia} combined with
hadronic asymmetry weights calculated from leading order
formulae, b) the polarized event generator
{\sc Sphinx} which treats the partonic helicity in the hard scattering and the initial state 
shower explicitly, and c)
parton generators which integrate next-to-leading order QCD cross sections using Monte Carlo
methods. Method~a) requires far less computer time than
method~b) but we find that it is too imprecise for precision studies. The lacking treatment
of partonic polarization in the initial state shower results in relative
deviations in the asymmetries of up to 20\% relative to {\sc Sphinx} (method~b)).
As was to be expected, the event generator
predictions for the absolute cross sections differ significantly from the 
next-to-leading order QCD calculations. But fortunately 
the asymmetries show a much better agreement: relative deviations
for prompt photon production for parameterizations with medium and large 
$\Delta g(x,\mu^2)$ barely exceed 10\%. After fine tuning the parameters of the
parton showers the same result can be obtained for jet production.
\end{abstract}
\pacs{}
%
%
%
%
%
%
%
%
\section{Introduction}
\label{section1}
One of the most remarkable features of high-energy hadron collisions  is
certainly their complexity. Particle multiplicities in single events
are frequently of the order ${\cal O}(10^2$-$10^3)$ and will continue
to rise with the advent of new colliders like {\sc Bnl-Rhic} and 
{\sc Cern-Lhc}. On the other hand, perturbative QCD (pQCD) can only deal with
a limited number of final-state partons so that we have to ask,
how these are related to experimentally measured distributions of hadrons and
leptons. This is where event generators (EGs) come to the rescue.

Heavy use of EGs is made during the whole life cycle of a 
high energy physics experiment for purposes such as estimating absolute production
rates, studying the background-to-signal ratio, studying the detector acceptance,
testing the reliability of event reconstruction during experimental analysis, etc.
For collisions of polarized hadrons this list must be extended to also include
studying the size of the hadronic double spin asymmetries of the signal and the
background, determining the sensitivity to the polarized parton distribution 
which is to be measured, calculating the {\em polarized acceptance}, etc. The last
point is especially important for fragmented detector systems with a small geometrical
coverage like {\sc Phenix} at {\sc Rhic} (see, e.g., \cite{martin3}). 
EGs for the simulation of collisions of 
longitudinally polarized particles must therefore include a correct treatment of 
particle helicity. 

In order to see how this goal can be achieved let us summarize the principle of
current EGs. The generation process is subdivided into several steps where the
first one deals with the hard partonic scattering. For nucleon-nucleon 
scattering, parton flavours 
and kinematics are selected according to the factorization formula for the
unpolarized cross section,
\begin{equation}
\label{eq:unpolfakt}
\DD \sigma^{(AB)}(P_A;P_B)=\sum_{a,b}\int\DD x_A\DD x_B 
f^{A}_{a}(x_A,\mu^2)
f^{B}_{b}(x_B,\mu^2)  \DD\hat
\sigma_{a,b}(x_AP_A,x_BP_B,\mu)\nonumber\,.
\end{equation}
A, B (a, b) are the involved hadrons (partons), $\mu$ denotes the factorization and 
renormalization scale and the partonic cross section $\DD\hat \sigma_{a,b}$ is evaluated to 
leading order (LO) in the strong coupling $\alpha_s(\mu)$. 
Please note, that the momenta of the colliding partons 
are so far parallel to the momenta of the colliding beams. 
Of course this picture is very simplified and is significantly altered by
radiative corrections which are specific to every hard partonic reaction. However,
certain parts of these radiative corrections, consisting of the collinear poles of each 
parton splitting, are universal and can be used to formulate a parton shower (PS)
algorithm which can be applied to every partonic subprocess \cite{sjoshower}.
In this way an approximation of the complete perturbative series is obtained which
becomes exact in the limit of vanishing decay angles and
fails for large angles. The parton showers can be somewhat arbitrarily subdivided
into an initial state shower (ISS) and a final state shower (FSS). The former generates
all particles which directly `connect' the two initial colliding partons  
with the ones that finally participate in the hard partonic reaction. It mimics
the DGLAP evolution of parton densities in normal perturbative calculations but there
is also an important difference:
each parton branching also involves transverse momentum which in general
leads to a Lorentz boost and rotation
of the partonic scattering system. Thus, the ISS changes the kinematics of the 
partons (for details see \cite{sjoshower}). Finally, the numerous partons
produced by the PS fragment into stable and unstable hadrons which are then
allowed to decay. Thus, EGs provide complete events which are in principle as detailed
as real events that could be measured with a perfect detector. In an ideal case, they
would also show the same fluctuation properties.

An approximate treatment of polarization can be achieved relatively easy and
has so far been used by the members of the {\sc Rhic} Spin Program \cite{rsc} to 
prepare the experiments \cite{jaffe4}:
\begin{itemize}
\item
Generate unpolarized events with the unpolarized EG {\sc Pythia} \cite{pythia}.  Extract
the kinematics  ($x_A$, $x_B$, $\hat s$, $\hat t$, $\hat u$, $\mu$) and the flavours of the 
hard partonic interaction. The partonic Mandelstam variables follow the usual definitions
\begin{equation}
\hat s=(x_AP_A+x_BP_B)^2\,,\quad 
\hat t=(x_AP_A-k_1)^2\,,\,\,{\rm and}\quad
\hat u=(x_AP_A-k_2)^2\,,
\end{equation}
where $k^{\mu}_{1,2}$ are the momenta of the outgoing partons.
\item
For each observable plot two histograms. The first histogram with the 
weight set to 1 gives the unpolarized cross section. For the second one the 
weight is taken to be the hadronic double spin asymmetry 
\begin{equation}
A_{LL}^{(AB)}(x_A,x_B,\hat s,\hat t,\hat u,\mu^2)=
\frac{\Delta f^A_a(x_A,\mu)}{f^A_a(x_A,\mu)}
\frac{\Delta f^B_b(x_B,\mu)}{f^B_b(x_B,\mu)}
\frac{\Delta\hat\sigma_{a,b}(\hat s,\hat t,\hat u,\mu^2)}{
\hat\sigma_{a,b}(\hat s,\hat t,\hat u,\mu^2)}\,.
\end{equation}
The formulae for the partonic double spin asymmetries 
$\Delta\hat\sigma_{a,b}/\hat\sigma_{a,b}$ for
the processes relevant for this work can be found in Table~\ref{tab1}.

\begin{table}[tbh]
\begin{center}
\begin{tabular}{cccl}
\\
{\sc Pythia} &  partonic & partonic asymmetry
$\frac{\Delta\hat\sigma}{\hat\sigma}$ & remark \\
 process number & reaction & &\\
\\
\hline
\hline
\\
11 & $qq'\rightarrow qq'$ & $\left(\hat s^2-\hat u^2\right)/\left(\hat s^2+\hat u^2\right)$ 
     \rule[0cm]{0cm}{0cm} & \\
   & $q\bar q'\rightarrow q\bar q'$ & $\left(\hat s^2-\hat u^2\right)/
      \left(\hat s^2+\hat u^2\right)$ 
     \rule[0.5cm]{0cm}{0cm}& \\
   & $q\bar q\rightarrow q\bar q$ & $\left[\hat s\left(\hat s^2-\hat u^2\right)
      +\frac{2}{3}{\cal I}\hat t\hat u^2/{\cal K}\right]/\left[
      \hat s\left(\hat s^2+\hat u^2\right)-\frac{2}{3}{\cal I}\hat t\hat u^2/{\cal K}\right]$
      \rule[0.5cm]{0cm}{0cm}& $\hat t$- and $\hat u$-channel only\\
   & $qq\rightarrow qq$ & $\left(\hat s^2-\hat u^2\right)/\left(\hat s^2+\hat u^2\right)$ 
      \rule[0.5cm]{0cm}{0cm}&  colour flow scenario~1\\
   & & $\left[\hat t\left(\hat s^2-\hat t^2\right)-\frac{2}{3}{\cal I}\hat s^2\hat u\right]/
      \left[\hat t\left(\hat s^2+\hat t^2\right)-\frac{2}{3}{\cal I}\hat s^2\hat u\right]$
      \rule[0.5cm]{0cm}{0cm}& colour flow scenario~2\\
12 & $q\bar q\rightarrow q\bar q$ & $-1$\rule[0.5cm]{0cm}{0cm}&  $\hat s$-channel only \\
   & \hspace{2mm}$q\bar q\rightarrow q'\bar q'$ & $-1$\rule[0.5cm]{0cm}{0cm}& \\
13 & $q\bar q\rightarrow gg$ & $-1$\rule[0.5cm]{0cm}{0cm} & \\
14 & $q\bar q\rightarrow g\gamma$ & $-1$\rule[0.5cm]{0cm}{0cm} & \\
18 & $q\bar q\rightarrow \gamma\gamma$ & $-1$\rule[0.5cm]{0cm}{0cm} & \\
28 & $qg \rightarrow qg$ & $-1$\rule[0.5cm]{0cm}{0cm} & colour flow scenario~1 \\
   & &\hspace{1.7ex}$1$\rule[0.5cm]{0cm}{0cm}& colour flow scenario~2 \\
29 & $qg \rightarrow q\gamma$ & $\left(\hat s^2-\hat u^2\right)/
      \left(\hat s^2+\hat u^2\right)$ & \\
53 & $gg\rightarrow q\bar q$ & $-1$\rule[0.5cm]{0cm}{0cm}& \\
68 & $gg\rightarrow gg$ & $-\left[\hat t^2+2\hat s\hat t\left(\hat s^2+\hat t^2\right)+
      3\hat s^2\hat t^2\right]/\left[
     \hat s^2+\hat t^2+2\hat s\hat t\left(\hat s^2+\hat t^2\right)+3\hat s^2\hat t^2\right]$
     \rule[0.5cm]{0cm}{0cm} & colour flow scenario~1\\
   & & $-\left[\hat u^2+2\hat s\hat u\left(\hat s^2+\hat u^2\right)+
     3\hat s^2\hat u^2\right]/\left[
     \hat s^2+\hat u^2+2\hat s\hat u\left(\hat s^2+\hat u^2\right)+3\hat s^2\hat u^2\right]$
     \rule[0.5cm]{0cm}{0cm} & colour flow scenario~2\\
   & & $2\left[\hat t\hat u\left(\hat t^2+\hat u^2\right)+
        3\hat t^2\hat u^2\right]/\left[\hat 
     t^2+\hat u^2+2\hat t\hat u\left(\hat t^2+\hat u^2\right)+3\hat t^2\hat u^2\right]$
     \rule[0.5cm]{0cm}{0cm} & colour flow scenario~3\\
\end{tabular}
\end{center}
\caption{\sf\label{tab1}
Partonic double spin asymmetries of all LO processes relevant to the production of
QCD jets and prompt photons.
The enumeration of color flow scenarios follows the values of the internal {\sc Pythia}
variable {\tt MINT(2)}~[6]. 
${\cal I}=1$, if the color interference terms are to be included, otherwise ${\cal I}=0$. 
${\cal K}$ is a $K$-factor with the default value ${\cal K}=1$.}   
\end{table}

\item
After the 
event generation is completed, divide the second histogram by the first one to obtain 
the hadronic double spin asymmetry. Finally, normalize the first histogram appropriately.
\end{itemize}

We shall call this prescription the method of asymmetry weights (MAW) in the remainder
of this paper. It is clearly only an approximation to the 'true answer' within the
EG model because it does not describe how parton showers depend on the parton helicities.

A simple way to overcome this deficit was presented by the authors of the 
EG {\sc Sphinx} \cite{sphinx} which is a polarized version of {\sc Pythia}. It uses
hadronic and partonic cross sections which depend on the helicities of the incoming
hadrons $H_{A,B}$ and partons $h_{a,b}$ 
and which are summed over the polarization of the outgoing particles.
The kinematics of the partonic reaction is therefore generated according to
\begin{eqnarray}
\DD \sigma^{(AB)}(P_A,H_A;P_B,H_B)&=&\sum_{a,h_a,b,h_b}\int\DD x_A\DD x_B 
f^{A,H_A}_{a,h_a}(x_A,\mu^2)
f^{B,H_B}_{b,h_b}(x_B,\mu^2) \DD\hat
\sigma_{ah_a,bh_b}(x_AP_A,x_BP_B,\mu^2)\,,\\
f^{A,H_A}_{a,h_a}(x_A,\mu^2)&=&\frac{1}{2}
\left[f^{A}_a(x_A,\mu^2)+\delta_{\epsilon(H_A),\epsilon(h_a)}\Delta f^{A}_a(x_A,\mu^2)
-\delta_{\epsilon(H_A),-\epsilon(h_a)}\Delta f^{A}_a(x_A,\mu^2)\right]\,,
\end{eqnarray}
where $\epsilon(x)$ is the sign function \cite{guellenstern}. 
Similarly, the helicity of all partons
participating in the ISS is chosen according to the helicity dependent DGLAP splitting 
functions
\begin{equation}
P_{bh_b,ah_a}(z) =
\frac{1}{2}\left[ P_{ba}(z)+\delta_{\epsilon(h_b),\epsilon(h_a)}\Delta 
P_{ba}(z)-\delta_{\epsilon(h_b),-\epsilon(h_a)}\Delta P_{ba}(z)
\right]
\end{equation}
instead of using the usual unpolarized $P_{ba}(z)$ as in {\sc Pythia}. Following this
procedure one directly gets the cross sections for different helicity configurations
of the colliding hadrons. They can be combined according to
\begin{equation}
A_{LL}^{(AB)}=\frac{\DD \sigma^{(AB)}(P_A,+;P_B,+)-\DD \sigma^{(AB)}(P_A,+;P_B,-)}{
\DD \sigma^{(AB)}(P_A,-;P_B,+)+\DD \sigma^{(AB)}(P_A,+;P_B,-)}
\end{equation}
to obtain the hadronic double spin asymmetry. It is clear, that one needs events as generated
by {\sc Sphinx} to test the experimental analysis software. {\sc Sphinx} has, however, a
severe disadvantage in 
comparison to the MAW. Longitudinal double spin asymmetries in $\vec p\vec p$-collisions are
frequently of the order of 1\% which means that $10^6$ {\em accepted} events
per bin are necessary to achieve a relative statistical Monte Carlo (MC) error of 10\%. 
The MAW needs much less
computation time since it calculates the asymmetry directly and not by computing the
relative difference between two similarly large numbers. 
Furthermore, it allows for the simultaneous usage of
multiple parameterizations of polarized parton distributions because they are 
irrelevant for the generation of the event itself. Depending on the size of the 
asymmetry computation times easily differ by a factor of the order ${\cal O}(10^2)$.
However, this speed advantage is useless if both methods yield largely different results
for the hadronic double spin asymmetries. So far, no systematic comparison 
using realistic experimental cuts for the relevant observables has been performed for
the {\sc Rhic} Spin Project or any other experiment according to the authors' knowledge. 
Since this question is so important we have done the necessary simulations and 
present our results in this paper.

We would also like to stress that rates and asymmetries calculated with {\sc Sphinx}
should not be regarded as the ultimate truth. EGs possess some serious 
shortcomings which their users should always be aware of. The foremost problem stems
from the fact that LO expressions for the cross sections and splitting functions
are used throughout. The resulting large scale dependences prohibit reliable predictions
of absolute cross sections or rates. Changes in some of the numerous 
necessary parameters of the PS model also may have quite strong effects on counting rates.
Being aware of this issue we should try to clarify if the polarized EGs nevertheless yield a good
description of nature. Unfortunately, this is currently impossible because
so far no $\vec p\vec p$-collisions have been studied. We can only compare
with the unpolarized results for the wide range of
available $pp$- and $p\bar p$-data but even this task is non-trivial without access
to the analysis software of the various collaborations. 

Since we are more interested in the polarized sector anyway, we decided to rather study
the uncertainty of theoretical predictions by comparing the EG results with next-to-leading
order (NLO) QCD calculations. In some sense, both methods are complementary because 
they take different parts of the perturbative series into account. Furthermore,
NLO predictions depend on a minimal set of parameters which includes only the
intrinsic QCD scale $\Lambda_{\rm QCD}$, the renormalization, factorization 
and in some cases the fragmentation scales. Usually this dependence is much smaller
than in LO calculations effectively giving NLO QCD predictive
power with regards to absolute rates. In some cases, agreement with experiment has
been quite spectacular. E.g., the NLO calculation of the transverse energy spectrum of
inclusive 1-jet production at {\sc Tevatron} at $\sqrt{S}=1.8$~GeV agrees very well with data
even though the unpolarized cross section varies over a range of six orders in magnitude 
\cite{cteq5}. Unfortunately, NLO calculations tend to get unrealiable if highly restrictive 
experimental acceptances are involved. We 
assume that NLO QCD will also do a good job for polarized
jet production at {\sc Rhic} which was shown to be very sensitive to the polarized gluon 
distribution $\Delta g(x,\mu^2)$ \cite{deflorian}, the prime target of most measurements 
at {\sc Rhic} and take the differences when compared to EG results as estimate for the
theoretical systematic error. 
Originally, prompt photon production has been promoted as the most
useful process for the measurement of $\Delta g(x,\mu^2)$
but currently the situation in the unpolarized
sector is far from satisfactory. Aside from the fact that NLO QCD cannot fit all
available data sets with one set of scales in the fixed-target energy range, there
also seem to be inconsistencies between some data sets \cite{aurenche}. 
Under very lucky circumstances these problems may happen to cancel when taking ratios of 
polarized and unpolarized cross sections but of course we should not count on it.
Whatever the solution to the current problems will be, NLO QCD will always be an integral
part of it. For this reason, we also decided to not only compare jet observables
but to also undertake a systematic comparison of prompt photon observables calculated
with EGs and NLO QCD. For this purpose we use the parton generators of
\cite{deflorian,frixione3} which use MC techniques to integrate NLO cross sections
and permit the implementation of simple cuts on the final state.

After discussing the necessary technical preliminaries we will present our
results for jet observables in section~2 whereas section~3 will deal
with prompt photon production only. The final section summarizes our
findings and contains the conclusions.
%
%
%
%
%
%
%
%
\section{Production of QCD-jets}
\label{section2}
The authors of Ref.~\cite{deflorian} have already studied 1-jet and 2-jet observables
in longitudinally polarized $\vec p\vec p$-collisions at the maximal {\sc Rhic} cm-energy of
$\sqrt{S}=500$~GeV using their NLO QCD parton generator. {\sc Star} is the only 
available detector system at {\sc Rhic} 
which is able to perform such measurements. After installation of the
endcap its acceptance region will cover $-1<\eta<2$ in pseudorapidity and $0<\phi<2\pi$
in azimuthal angle. The main results of \cite{deflorian} are that the perturbative
series of the polarized cross section is under control, that parameterizations
with different $\Delta g(x,\mu^2)$ yield very distinctive asymmetries and that
the statistical experimental error will be sufficiently small to allow for a 
precise measurement from which the polarized gluon distribution can be determined.
Therefore, \cite{deflorian} is an ideal basis for our studies and we can use 
the same observables without any worries about perturbative stability.

Before we give a detailed description of them let us make a few remarks on the choice of
the parton distributions and the QCD running coupling constant. Nowadays, polarized inclusive
lepton-nucleon scattering still has the by far largest impact on the determination
of the polarized parton distributions. Unfortunately, the current data are not
precise enough and cover a too small $x$-$Q^2$-range to give any useful restriction
on $\Delta g(x,\mu^2)$.  As a consequence, LO and NLO fits based on the same
assumptions and restrictions may give widely different results for the 
polarized gluon distribution, i.e. the ratio 
\begin{equation} 
\label{eq:kg}
K_{\Delta g}(x,\mu^2)= \frac{\Delta g^{\rm NLO}(x,\mu^2)}{\Delta g^{\rm LO}(x,\mu^2)}
\end{equation}
can deviate strongly from 1. According to Fig.~\ref{fig1} this is indeed the case for two of 
the three sets of polarized parton distributions we use in our studies. Although
the EGs are only based on LO QCD we would therefore be ill advised to use them in
combination with LO parton distributions. Any agreement with NLO QCD results would
be purely accidental, especially in view of the sensitivity of jet production towards
$\Delta g(x,\mu^2)$. Accordingly, we also use the NLO expression 
for the strong coupling $\alpha_s(\mu^2)$ throughout to ensure that the EG without 
PS and the lowest order perturbative calculation will give the same results. 
The unpolarized GRV \cite{grv} as well as the polarized GRSV \cite{grsv} and DSS \cite{dss}
parameterizations we use are already quite old. Nevertheless, they still 
fulfill all our needs since we
are interested in comparing different methods of calculation rather than making
precise predictions. It is more important here to ensure that for each polarized
set we use the correct unpolarized set of distributions which in all cases is GRV. 
Otherwise {\sc Sphinx} will not give correct results because it makes use of the
unpolarized and polarized set simultaneously, so that only one value can be given for 
$\Lambda_{\rm QCD}$. The three selected sets of $\Delta f(x,\mu^2)$ pretty much cover the
whole range of possible values for $\Delta g(x,\mu^2)$ with GRSV $g_{\rm max}$ giving
the largest and DSS3 the smallest values. In all cases the scales were set equal to
the average transverse momentum of the partons leaving the hard partonic interaction:
\begin{equation}
\label{eq:scalechoice}
\mu=\frac{1}{n}\sum_{i=1}^n|k_{T,i}|\,.
\end{equation}

\subsection{The observables}
We already mentioned that the number of final-state partons is much larger 
in the EG model than in perturbative calculations even though both
intend to describe the same physics. For our studies we should therefore pick a
jet-clustering algorithm which depends only weakly on particle multiplicities.
The Ellis-Soper $k_T$-algorithm \cite{es} with the preferred choice of $R=1$
suits our needs. For each event it gives the number of
reconstructed jets as well as their transverse energies $E_T$, pseudorapidities $\eta$
and azimuthal direction $\phi$. From these we construct four observables\footnote{
We actually studied more observables but they did not yield any new insights.} 
which are defined as follows:
\begin{itemize}
\item 
the $E_T$-dependent 1-jet inclusive cross section
$\DD(\Delta)\sigma/\DD E_T$, 14~GeV$<E_T<50$~GeV;
\item 
the $\eta$-dependent 1-jet inclusive cross section
$\DD(\Delta)\sigma/\DD\eta$, $0<\eta<2$, $E_T>15$~GeV.
Since the cross section is symmetric in $\eta$ we restrict ourselves to
the region of positive pseudorapidity.
\end{itemize}
For all 2-jet inclusive observables we denote the hardest and second hardest jet by
$J_1$ and $J_2$ and demand
\begin{equation}
\label{eq:2jets}
E_{T,J_1}>15\,\,{\rm GeV}\,,\quad E_{T,J_2}>10\,\,{\rm GeV}\,,\quad 
|\eta_{J_1}|<1\,,\quad |\eta_{J_2}|<1\,.
\end{equation}
The asymmetric cut on the transverse energies reduces contributions from
configurations in which the two leading jets are exactly back-to-back and
which cannot be calculated with finite order perturbation theory \cite{deflorian}.
All events fulfilling the requirement (\ref{eq:2jets}) are considered for
\begin{itemize}
\item the distribution of the rapidity difference 
$\DD(\Delta)\sigma/\DD\Delta\eta\,$, $0<\Delta\eta<2\,$,
$\Delta\eta=\eta_{J_1}-\eta_{J_2}\,$;
\item
the distribution of the invariant jet mass $M_{JJ}$ defined by
$\DD(\Delta)\sigma/\DD M_{JJ}\,$, $M_{JJ}<100$~GeV,
$M_{JJ}^2=(p_{J_1}+p_{J_2})^2$.
\end{itemize}

\subsection{Comparison of the EG methods}
Event generators are very complex programs which often consist of several ten thousand
lines of code and are equipped with a large number of options. In order to avoid 
errors we should try to reuse as much program code as possible for our comparisons
of EG methods and NLO calculations. This can be easily achieved, 
because {\sc Sphinx} can also be operated in an unpolarized mode, in which it corresponds to
{\sc Pythia}~5.6 and only needs to be extended with a function that calculates the 
asymmetry weights of Table~\ref{tab1}. 
Furthermore, the EG and the NLO parton generators use the
same code to evaluate the parton distributions and the strong coupling constant.

The absolute statistical MC error of the asymmetries calculated with {\sc Sphinx}
is only determined by the number of accepted events in a certain bin and is therefore 
independent of the actual size of the asymmetries. In order to maximize the precision
and reduce the required computer time we should use parton densities which yield
large asymmetries, i.e. GRSV~$g_{\rm max}$ is the preferred set.
The artifical set $\Delta f(x,\mu^2)=f(x,\mu^2)$ generally yields even larger asymmetries
but cannot be used within the PS model since it is inconsistent with DGLAP evolution.
By running {\sc Sphinx} and {\sc Pythia} with PS switched {\em off}, we made sure that the
implementation of the asymmetry weights is indeed correct. The relative deviations
between asymmetries calculated with {\sc Sphinx} and the MAW did not exceed 2\% and were 
within the statistical MC accuracy. We want to stress, that during all 
EG runs the hadronization and intrinsic partonic transverse momentum were switched off
so that the results can be better compared to NLO calculations.

Fig.~\ref{fig2} shows the results for the (un)polarized cross sections, asymmetries
and the relative deviations
\begin{equation}
\label{eq:reldiff1}
R_{\sigma}= \frac{\DD\sigma_{\rm MAW}-\DD\sigma_{\rm\sc Sphinx}}{
\DD\sigma_{\rm MAW}}\,,\quad
R_{A}= \frac{A_{LL,\rm MAW}-A_{LL,\rm\sc Sphinx}}{A_{LL,\rm MAW}}
\end{equation}
between the MAW ($8\cdot 10^6$ generated events) and {\sc Sphinx} 
($4\cdot 10^7$ generated events) with the PS switched {\em on}. Let us concentrate on the
right column, in which the full black line represents $R_{\sigma}$. In all cases it 
is either compatible with or very close to zero so that we proved once more that the spin
averaging of {\sc Sphinx} indeed works. The dashed crosses represent $R_A$ calculated
with {\em polarized treatment switched on} in the ISS of {\sc Sphinx}. For all 
four observables systematic deviations show up. The asymmetries for the two 1-jet 
observables calculated with the MAW are generally about 5\% larger than the ones which
are calculated with {\sc Sphinx}. The differences are especially strong in the
large $\eta$-region, where they reach 20\%. However, these findings cannot
be generalized to all observables. For the $\Delta\eta$-distributions and the
region of small invariant mass the situation is reversed. Within the limited accumulated
MC statistics, both methods seem to yield the same asymmetries for large $M_{JJ}$ but even
here we can note a tendency for systematic differences towards positive $R_A$.
In order to pin down the origin of the large deviations we also carried out 
an analysis of events generated by {\sc Sphinx} for which the 
{\em polarized treatment in the ISS was switched off} (dotted crosses in Fig.~\ref{fig2}).
From the fact, that $R_A(\eta)$ is now compatible with zero for large pseudorapidity
we conclude, that the largest part of the deviations is indeed caused by the
missing treatment of parton helicity within the MAW approach. However this cannot be
the full story because $R_A$ is not compatible with zero for all other observables.
But keeping in mind that the ISS changes the kinematics of the hard partonic interaction,
that within the MAW the ISS is applied to unpolarized events whereas within {\sc Sphinx}
it is applied to polarized events with different configurations of proton helicities,
it is clear that the ISS can slightly modify the asymmetry even though polarized treatment
is switched off. Unfortunately we were unable to check if the same results also
hold true for the two other sets of polarized parton distributions. As we will see in 
Fig.~\ref{fig4}, the set GRSV~std. (DSS3) yields asymmetries which are smaller by 
a factor of five (twenty) so that we would have had to generate $1\cdot 10^9$ 
($1.6\cdot10^{10}$) events with {\sc Sphinx} to obtain the same relative statistical 
accuracy. This was simply beyond the means of our computer equipment.

\subsection{Comparison of the MAW with NLO QCD predictions}
In view of the fact that parton helicity plays a significant role in the ISS we would
have liked to compare {\sc Sphinx} with NLO QCD. Unfortunately, this was impossible
for the same exact reasons as were given at the end of the previous paragraph.
We resort to comparing the results of the MAW with NLO QCD predictions so that
several different parameterizations of polarized parton densities can be used.

In order to find a common basis we again start out by comparing Born-results, i.e.
observables calculated with LO partonic cross sections, NLO parton distribution,
and a NLO expression for the running strong coupling. For this purpose, Fig.~\ref{fig3}
shows the relative deviations of the unpolarized cross sections and the asymmetries 
defined by
\begin{equation}
\label{eq:reldiff2}
R_{\sigma}\equiv \frac{\DD\sigma_{\rm (N)LO\,\, QCD}-\DD\sigma_{\rm MAW}}{
\DD\sigma_{\rm (N)LO\,\, QCD}}\,,\quad R_{A}\equiv \frac{A_{LL,\rm (N)LO\,\,QCD}-
A_{LL,\rm MAW}}{A_{LL,\rm (N)LO\,\, QCD}}\,.
\end{equation}
Only the error bars of $R_{\sigma}$ are plotted for the sake of clarity, as the 
error bars of $R_A$ can be estimated from the statistical fluctuation between
two neighbouring bins. The unpolarized cross sections and the asymmetries for
both GRSV sets mainly show a satisfactory, but not perfect agreement for the bins 
with high statistics. This is unfortunately not the case for the DSS3 set --
even for bins with good statistics deviations of 5\% show up although both
programs use the same routines to evaluate the parton distributions, as was
already mentioned before. The reason for these problems is still under
investigation but we assume, that it is most likely caused by the different
treatment of colour interference terms (c.f. \cite{pythiamanual}) and the
fact, that the relatively small asymmetries of the large number of contributing
partonic processes undergo a delicate cancellation process.
We will soon see that PS or NLO corrections to the asymmetries will be by far 
stronger so that we can effectively neglect the just mentioned problems.

\begin{table}
\begin{center}
\begin{tabular}{llcc}
\\
{\sc Pythia} parameter & description & default value & optimized value \\
\\
\hline
\hline
\\
{\tt PARP(64)} & argument of $\alpha_s$ for each parton splitting divided by $k_{\bot}^2$  & 1 & 2.5 \\
{\tt PARP(65)} & minimal energy of an emitted parton         & 2~GeV & 3~GeV \\
{\tt PARP(67)} & maximal scale of the ISS divided by $\mu$   & 4     & 3     \\
\\
\end{tabular}
\end{center}
\caption{\sf\label{tab2}
Modified parameters of the ISS which were used for the calculations of Fig.~\ref{fig5}.
}
\end{table}

In an ideal case, the results of NLO QCD and the MAW would just agree. This is
of course very unlikely.  Just by considering the scale dependence
of the observables, which is much larger for the EG, we see that agreement
can only be achieved for one special scale or even not at all, because the
results might not only differ in their normalization but also in their shape.
So it is not surprising, that the cross sections calculated with both methods
deviate by up to 40\% when using the standard choice of scales and the
standard PS parameters (see Fig~\ref{fig4}). However, the objects we are
mainly interested in, namely the asymmetries, show a much better agreement. 
$R_A$ seldomly exceeds 20\% for any of the three polarized sets and the
agreement is about twice as `good' for $A_{LL}(M_{JJ})$.

The PS needs a number of parameters which can be freely varied within
certain limits and have to be determined by fitting to data. We picked three
of of them, listed in Table~\ref{tab2}, and tried to refit them using the
NLO QCD calculation as `data set'. Due to the long necessary computation
time an automated fitting procedure is unpractical and we resort
to manually testing a few values. The $E_T$- and $\eta$-dependent observables
are the most inclusive ones so that their normalization should be fixed first, aside from 
improving the agreement of the asymmetries. 
Accordingly, the absolute rates predicted by the EG have to be reduced which can be
achieved by reducing the activity of the ISS. Table~\ref{tab2} shows the
default as well as the fitted values of the three parameters. Since the 
fitting was done manually with a very limited number of different values
it is unlikely that the optimal set has been found. Nevertheless, a look at
Fig.~\ref{fig5} shows that it completely fulfills its purpose: the 
unpolarized cross sections for the 1-jet inclusive observables as well as
the asymmetries for the two GRSV sets now agree at a 10\% level whereas
the situation for the set DSS3 is not quite as good.
On the other hand, the agreement for the 2-jet inclusive rates got much
worse. 

In order to get a feeling for the size of the relative deviations we study
the scale dependence of the NLO predictions. In Fig.~\ref{fig6} 
the standard choice of the scales (\ref{eq:scalechoice}) is varied by a factor
of 0.5 and 2, resp., which results in the bands of cross section 
displayed in the left column of the figure. By dividing the minimal (maximal)
polarized cross section by the maximal (minimal) unpolarized cross section
we obtain the limits on the bands of asymmetries, which are shown in the
right column. According to the figure, the optimized EG asymmetries are 
always within the bands. Aside from the normalization of the 2-jet inclusive
cross sections, the EG results are therefore compatible with NLO QCD.

%
%
%
%
%
%
%
%
\section{Prompt photon production}
\label{section3}
Let us now turn to the production of prompt photons in $\vec p\vec p$-collisions.
This process receives a lot of interest mainly due to fact that the quark-gluon fusion 
$qg\rightarrow q\gamma$ strongly dominates at LO
for medium sized and large $\Delta g(x,\mu^2)$. Before the trouble with the
fixed-target data emerged, the analysis therefore seemed to be especially easy and clean.
At the large cm-energies accessible at {\sc Rhic} another problem arises: the majority
of photons are actually not produced in the hard partonic interaction but 
through pion-decay into two photons. These fragmentation photons can luckily 
be completely filtered out by an isolation prescription defined in Ref.~\cite{gammaiso}.
Unless the usual cone-isolation prescription, which limits the hadronic energy
contained in a cone drawn around the photon direction in $\eta$-$\phi$-space with
opening angle $\delta_0$, Frixione's prescription allows less hadronic energy the
closer the hadrons are to the photon and no hadronic energy for exactly parallel hadrons.
Since fragmentation is an exactly collinear process in QCD, this prescription effectively
eliminates all fragmentation photons while at the same time the cancellation of 
soft and collinear divergences stays untouched. For our studies we choose $\delta_0=0.7$.

\subsection{The observables}
Common experience with polarized prompt photon calculations shows that asymmetries
tend to become smaller when the cm-energy of the experiment is increased because
smaller $x$-values are probed. For our event generator studies we therefore choose
$\sqrt{S}=200$~GeV instead of the maximal {\sc Rhic} cm-energy which is beneficial for the
precision of all results calculated with {\sc Sphinx}. Our choice of observables
follows closely the NLO QCD study of Ref.~\cite{frixione3} for the same reasons that were 
given in section~\ref{section2}. The cuts were inspired by the geometrical coverage
of {\sc Star}.
Frixione's isolation criterion in combination with the ES jet algorithm yields
the transverse momentum, rapidity and azimuthal angle of the photon and a certain
number of jets for all events which pass the isolation cut. These are then used to
calculate two inclusive observables:
\begin{itemize}
\item
the $p_{T,\gamma}$-dependent cross section
$\DD(\Delta)\sigma/\DD p_{T,\gamma}$, 10~GeV$<p_{T,\gamma}<$50~GeV, $|\eta_{\gamma}|<0.5$;
\item
the $\eta_{\gamma}$-dependent cross section
$\DD(\Delta)\sigma/\DD \eta_{\gamma}$, $0<\eta_{\gamma}<2$, $p_{T,\gamma}>10$~GeV.
Due to symmetry we again restrict ourselves to the positive $\eta$-region.
\end{itemize}
Inclusive prompt photon observables have the advantage that counting rates are
quite large. On the other hand, the kinematics of the hard partonic reaction 
can of course not be reconstructed. Thus, there is no one-to-one correspondence 
between the measured transverse momentum of the photon and the probed gluonic $x$
even in a LO analysis. This problem can be fixed if only events with one reconstructed
jet are considered. Additionally, we require
\begin{eqnarray}
\label{eq:gjcut}
p_{T,\gamma}>10~{\rm GeV}\,,\quad E_{T,J}>11~{\rm GeV}\,,\quad -1<\eta_{\gamma,J}<2\,,
\quad |\Delta\phi|=|\phi_J-\phi_{\gamma}|>\frac{\pi}{2}
\end{eqnarray}
which effectively means that the jet must be located in the hemisphere opposite to
the photon. Events with multiple jets are not considered here. Their 
simulation would require a NLO parton generator with so far unknown matrix elements
of order ${\cal O}(\alpha\alpha_s^3)$ or higher. 
All events that pass the criteria (\ref{eq:gjcut})
are considered for two $\gamma J$-observables:
\begin{itemize}
\item
the distribution of the rapidity difference
$\DD(\Delta)\sigma/\DD\Delta\eta$, $0<\Delta\eta<2.4$, $\Delta\eta=\eta_J-\eta_{\gamma}$;
\item
the distribution of the invariant mass of the $\gamma J$-pair
$\DD(\Delta)\sigma/\DD M_{\gamma J}$, $M_{\gamma J}<100$~GeV, 
$M_{\gamma J}^2=(p_{\gamma}+p_{J})^2$.
\end{itemize}

\subsection{Comparison of the EG methods}
For the comparison of the EG methods we unfortunately had to restrict ourselves
again to the set GRSV $g_{\rm max}$. The mandatory test of the implementation
of the MAW did not yield any surprise. The asymmetries calculated from
Born-cross sections with both methods agreed within the relative statistical 
accuracy of a few per cent. 

Before discussing the full results including parton showers we should be careful
to check if Frixione's isolation method also works for the string fragmentation model used
in {\sc Pythia}. Hadronization is not an exactly collinear process here, so that
some fragmentation photons might actually fulfill the isolation criterion. 
This is indeed the case, but only for very few events so that we do not need to worry about 
them and can safely keep fragmentation switched off. Unlike for NLO QCD
calculations, where all perturbatively produced photons originate directly
in the hard interaction, we need to make a further distinction for 
perturbatively produced photons in the EG model. Aside from `direct' photons, which
are produced in the LO hard processes $qg\rightarrow q\gamma$ and 
$q\bar q\rightarrow g\gamma$, there is also the possibility that the photon is produced by 
emission from a quark line in the PS `before' or `after' one
of the purely strong interacting reactions $qq\rightarrow qq$, $qg\rightarrow qg$,
$gg\rightarrow gg$, etc. occurs. We denote this second source of
photons as `bremsstrahlung' photons. Depending on their transverse momentum  they make up 
20-40\% of all isolated photons at $\sqrt{S}=200$~GeV according to Fig.~\ref{fig7}.
The simulation of the bremsstrahlung contribution is very tedious
because under normal circumstances only every thousandth event 
contains an isolated photon. This very low efficiency can be somewhat improved
by multiplying the value of $\alpha$ by a small factor (e.g. 3) so that
the ratio of QED- to QCD-emissions becomes larger without changing the event
characteristics. But even then 
huge numbers of events are needed for decent statistical accuracies, especially 
when using {\sc Sphinx}. We therefore only present the comparison 
of 'direct' photons in Fig.~\ref{fig8}, for which we produced $1.7\cdot 10^8$ events. 
We proceed along the same lines as for the jet studies in the last section.
At first, the correct treatment of particle helicity in the ISS of {\sc Sphinx} is 
switched on resulting in the dashed crosses in the right column. Then the ISS is
operated in unpolarized mode resulting in the dotted crosses for $R_A$.
The overall picture is very similar to Fig.~\ref{fig2} but less pronounced.
Especially the $p_{T,\gamma}$- and $\Delta\eta$-dependent asymmetries  show
relative deviations of up to 10\% which are again due to the missing treatment
of particle helicity in the ISS within the MAW. 
Keeping in mind that we do not know if the inclusion of bremsstrahlung photons would 
change the overall picture, we can state that the usage of the MAW
for prompt photon simulations {\em appears} to be better justified than for jet physics.
We generated an additional $2.8\cdot 10^8$ events to study the bremsstrahlung contribution
but unfortunately the insufficient
 statistical accuracy did not allow us to draw any conclusions.

\subsection{Comparison of the MAW with NLO QCD predictions}
Before we proceed to the comparison with NLO QCD we should again try to establish
a common basis for both methods by checking the Born-level results. Our tests showed
that cross sections and asymmetries calculated with the MAW and the parton generator
agreed on a 1\% level so that everything is in perfect order. This is a strong 
indication for the correctness of our explanation for the deviations seen in 
Fig.~\ref{fig3} because colour interference does not exist in LO prompt photon 
production (see also Table~\ref{tab1}). 

In Fig.~\ref{fig9} the full EG results, i.e. including PS and including QED-bremsstrahlung,
are compared with the NLO QCD predictions. Please note, that an overall $K$-factor
of $K=1.5$ has been applied to all cross sections calculated with the MAW to bring its
rates better in line with the NLO predictions. Of course this correction is far
from being ideal but it can be justified by the large scale dependence
of the EG results. Indeed, it is common practice to more or less trust the shape of 
cross sections predicted by an EG and to adjust the normalization within certain limits.
A slightly larger $K=1.67$ would have led to a very good agreement for
$\DD\sigma/\DD\Delta\eta$ and $\DD\sigma/\DD M_{\gamma J}$ but for the price of larger
discrepancies between the inclusive cross sections whose slopes are already way off. 
With the exception of the critical set DSS3, the asymmetries surprisingly deviate only by 
5-10\% on average! Due to the small gluon parameterization of DSS3, 
resulting in the small asymmetries of Fig.~\ref{fig9},
the quark-gluon fusion process does not dominate as strongly and other processes become
increasingly relevant. The situation becomes a lot more complicated with the effect that
the discrepancies between asymmetries reach up to 25\%. 
Unfortunately, the very long computation times prohibited any attempt 
to improve the agreement by repeating the fit of the shower parameters discussed in
section~\ref{section2}.

This leaves us with the study of the scale dependence of the NLO predictions presented
in Fig.~\ref{fig10}. It turns out to be much larger than for jet production so that
the different bands of asymmetries partly overlap. With the exception of two
bins even the sizable deviations of DSS3 are within the limits. The asymmetries for the
two GRSV parameterizations are located near the center of the corresponding bands.
At least in this restricted sense, the asymmetries calculated with the MAW and NLO QCD
are consistent with each other.

\section{Resume}
Our detailed comparisons between {\sc Sphinx} and the method of asymmetry weights
show, that the correct description of particle polarization in the initial state shower
generally has to be implemented into any polarized event generator if one requires
a precision of order 10\%. For some of the
studied observables we find even differences as large as 20\% between the MAW and 
{\sc Sphinx}. The most drastic deviations arise for the region
of large pseudorapidity ($\eta\approx 2$) of 1-jet inclusive observables. Luckily this
region of phase-space will be irrelevant for the experiments at {\sc Rhic}. The
geometrical coverage of {\sc Star} only reaches out to $\eta=2$ so that under 
normal circumstances the maximal pseudorapidity for the jet-axis will be $\eta\approx 1$.
So we are left with maximal deviations of approximately 10\%, which are however
still substantial.
In view of the tremendous advantage in speed of the method of asymmetry weights
we therefore recommend the following compromise: in a first step, all studies should
be performed using the method of asymmetry weights. Once the optimal energies, 
experimental cuts, etc. have been found, {\sc Sphinx} should be used to check the results.

The comparisons between event generator and NLO QCD predictions are 
encouraging despite of the substantial discrepancies in the unpolarized sector.
The longitudinal double spin asymmetries are the objects we are mainly interested in and
the agreement is much better here. With very few exceptions, the deviations between
the method of asymmetry weights and NLO~QCD are smaller than the uncertainties 
of the NLO QCD predictions due to scale dependence. Both methods yield
consistent results in this restricted sense. During the starting phase of the 
{\sc Rhic} spin project these theoretical uncertainties on the asymmetries 
of approximately 10\% are probably tolerable.
However, in the long run more precise event generators will be needed. Their
development can in our opinion not be achieved without substantial conceptual progress
in event generator techniques. Most likely a fusion of NLO matrix elements with the 
parton shower methods is needed.
The development of such a NLO EG has been a hot topic of scientific debate 
\cite{nloeg} for several years already due to its urgency. So far, no working
solution is available. In view of the substantial costs of the {\sc Rhic} spin program 
and the scientific importance of a polarized NLO event generator it would 
seem advisable to us to also invest the necessary manpower and funds into the development of 
such a code, which, however, does not seem to happen.
Thus, we might soon be in the situation that wonderfully rich data is available but 
cannot be appropriately analyzed.

\section*{Acknowledgments}
This work was supported by the BMBF and the ``Deutsche Forschungsgemeinschaft''.
We thank the authors of Refs.~\cite{deflorian,frixione3} for making their 
parton generator codes available to us. O.~Martin is grateful to N.~Saito for
the invitation to the RIKEN-BNL research center workshop ``Event Generator for
Spin Physics'' and thanks A.~Kirchner, L.~Niedermeier, S.~Sch\"afer, E.~Stein and
M.~Stratmann for lively discussions during our weekly meetings.

%
%
%
%
%
%

%
%
%
%
%
%
\newpage
\section*{Figure captions}
\begin{figure}
\begin{center}
\epsfig{file=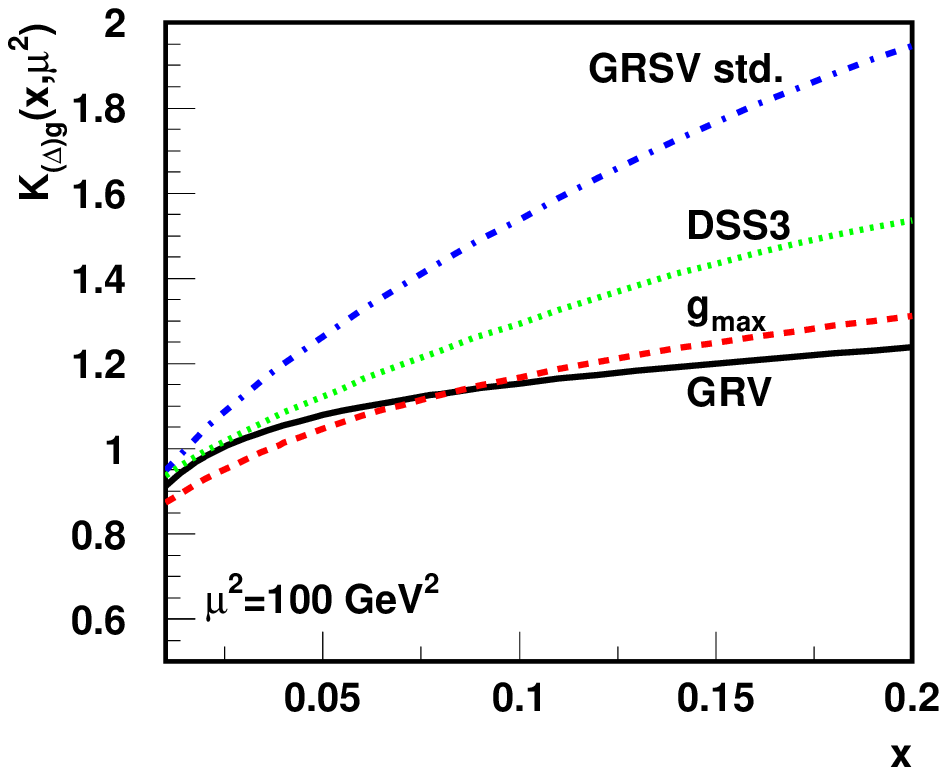,width=12cm}
\end{center}
\caption{\label{fig1}}
\end{figure}

\begin{figure}
\begin{center}
\epsfig{file=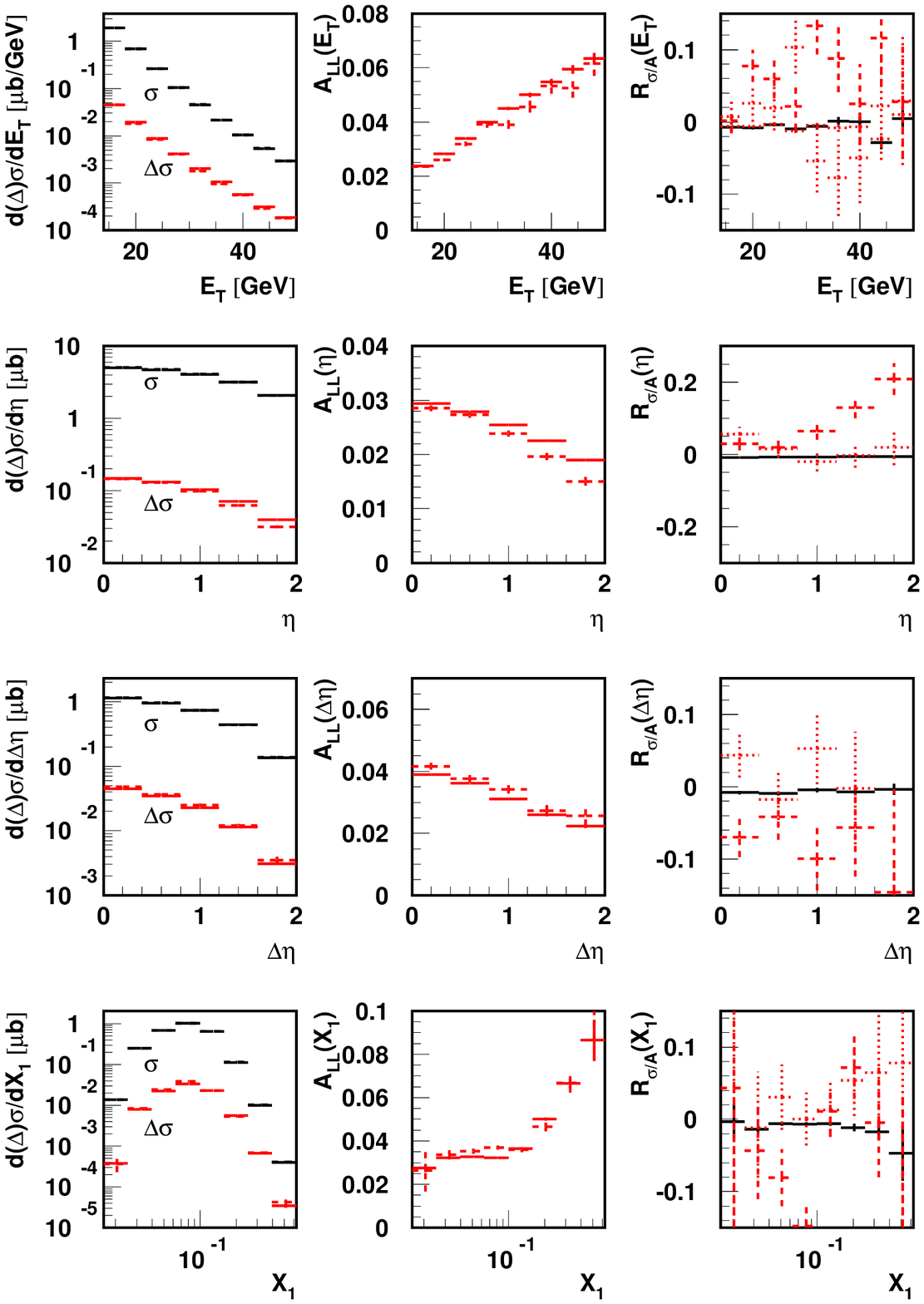,width=16cm}
\end{center}
\caption{\label{fig2}}
\end{figure}

\begin{figure}
\begin{center}
\epsfig{file=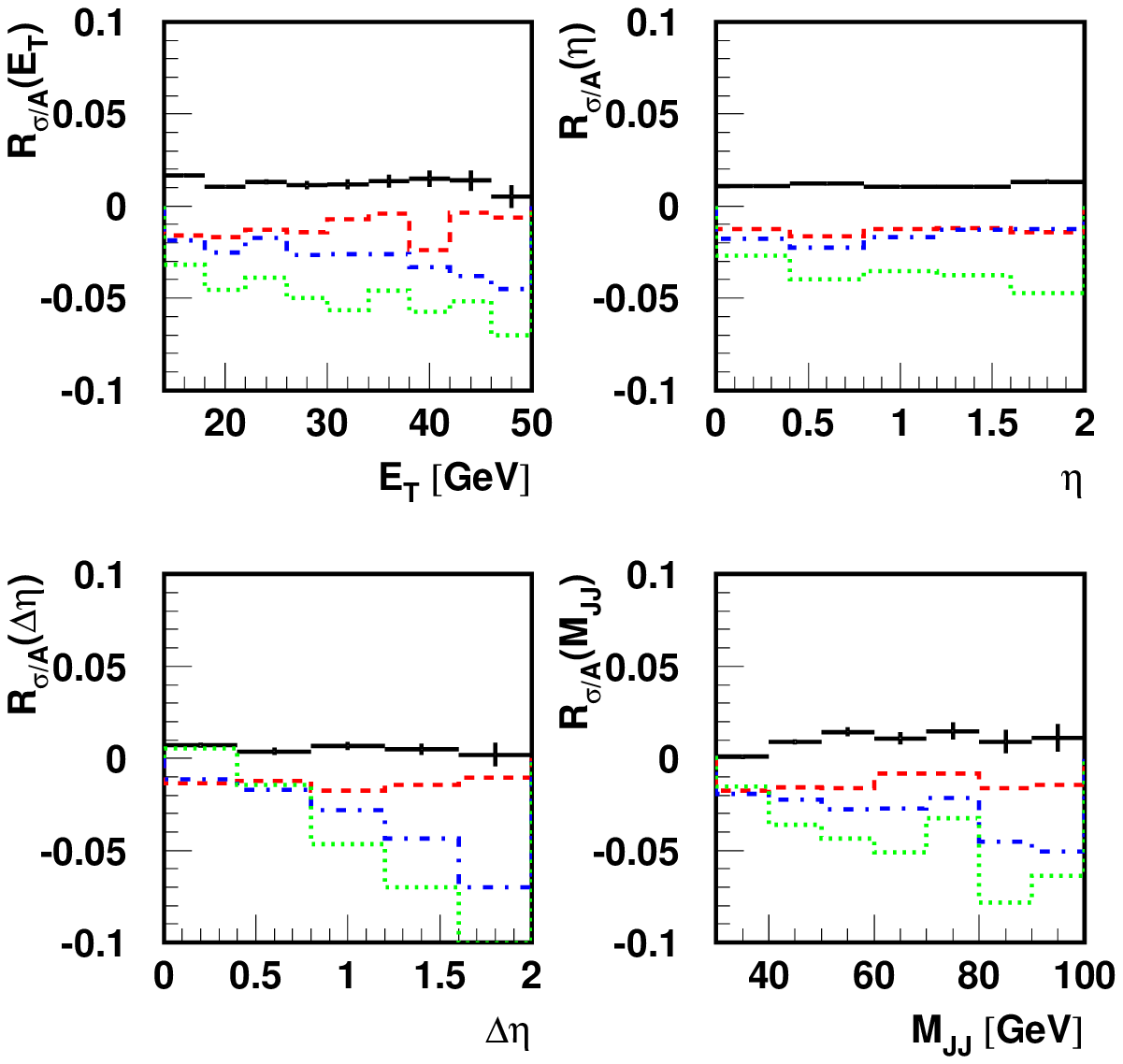,width=15cm}
\caption{\label{fig3}}
\end{center}
\end{figure}

\begin{figure}
\epsfig{file=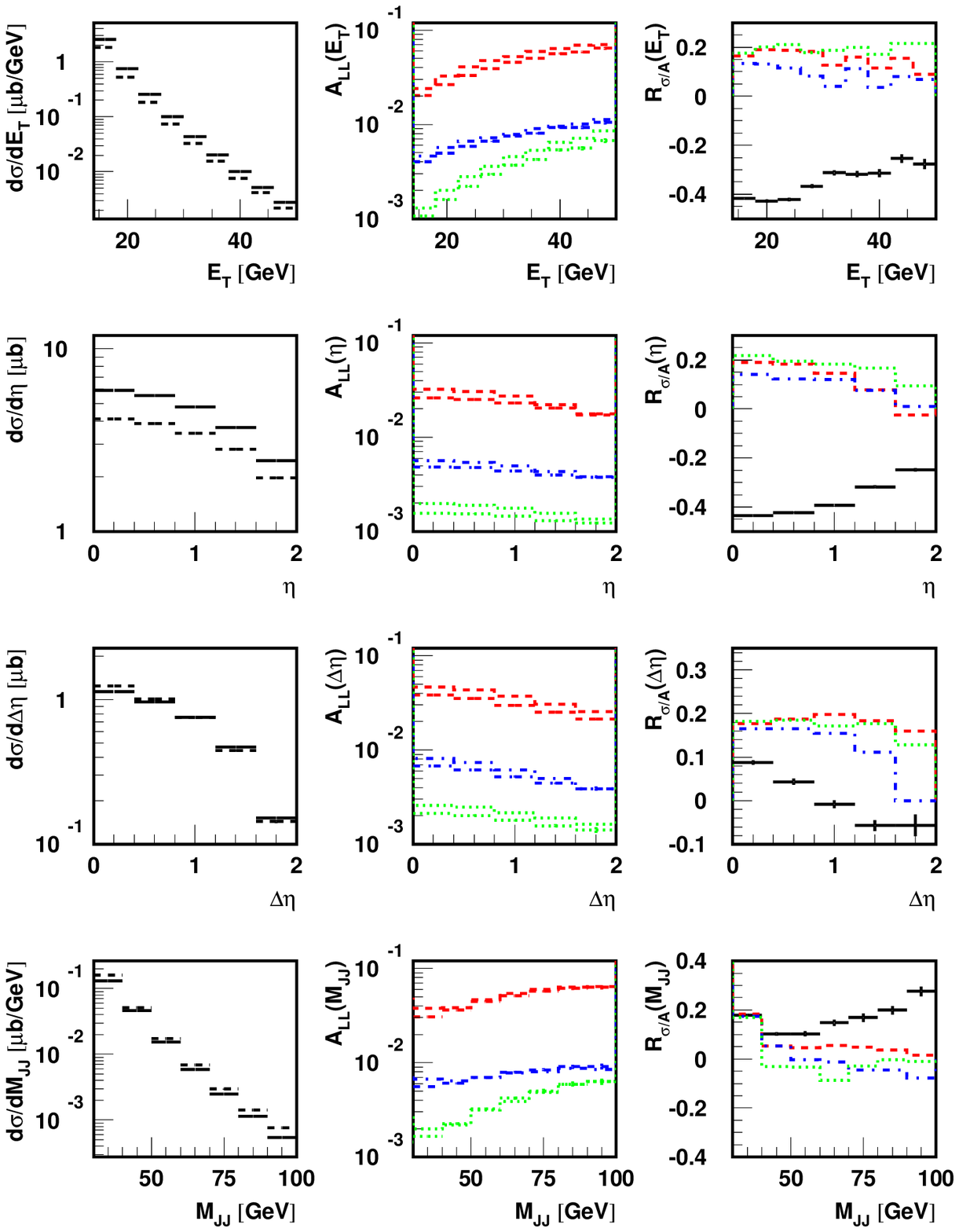,width=17cm}
\caption{\label{fig4}}
\end{figure}

\begin{figure}
\epsfig{file=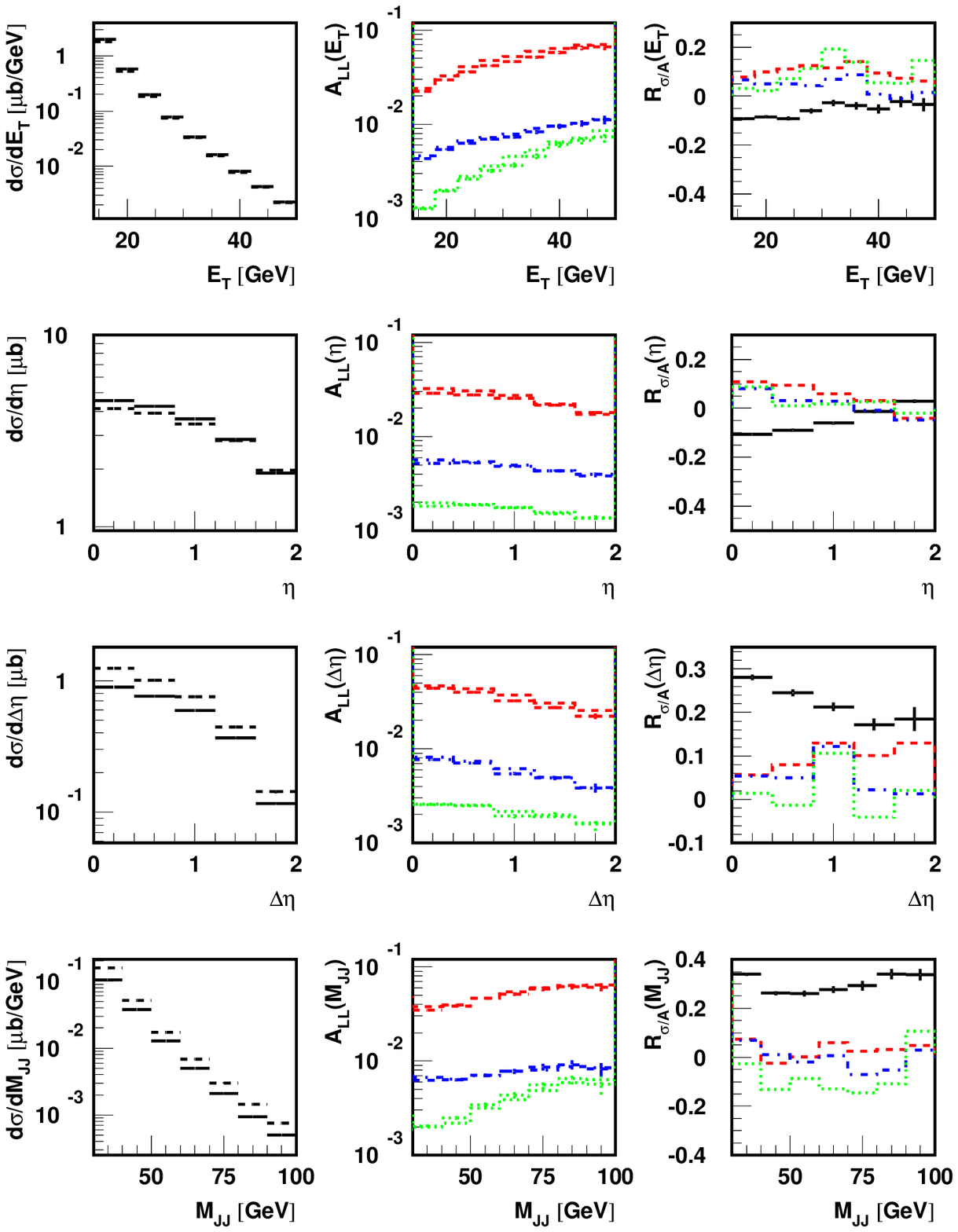,width=17cm}
\caption{\label{fig5}}
\end{figure}

\begin{figure}[p] 
\begin{center}
\epsfig{file=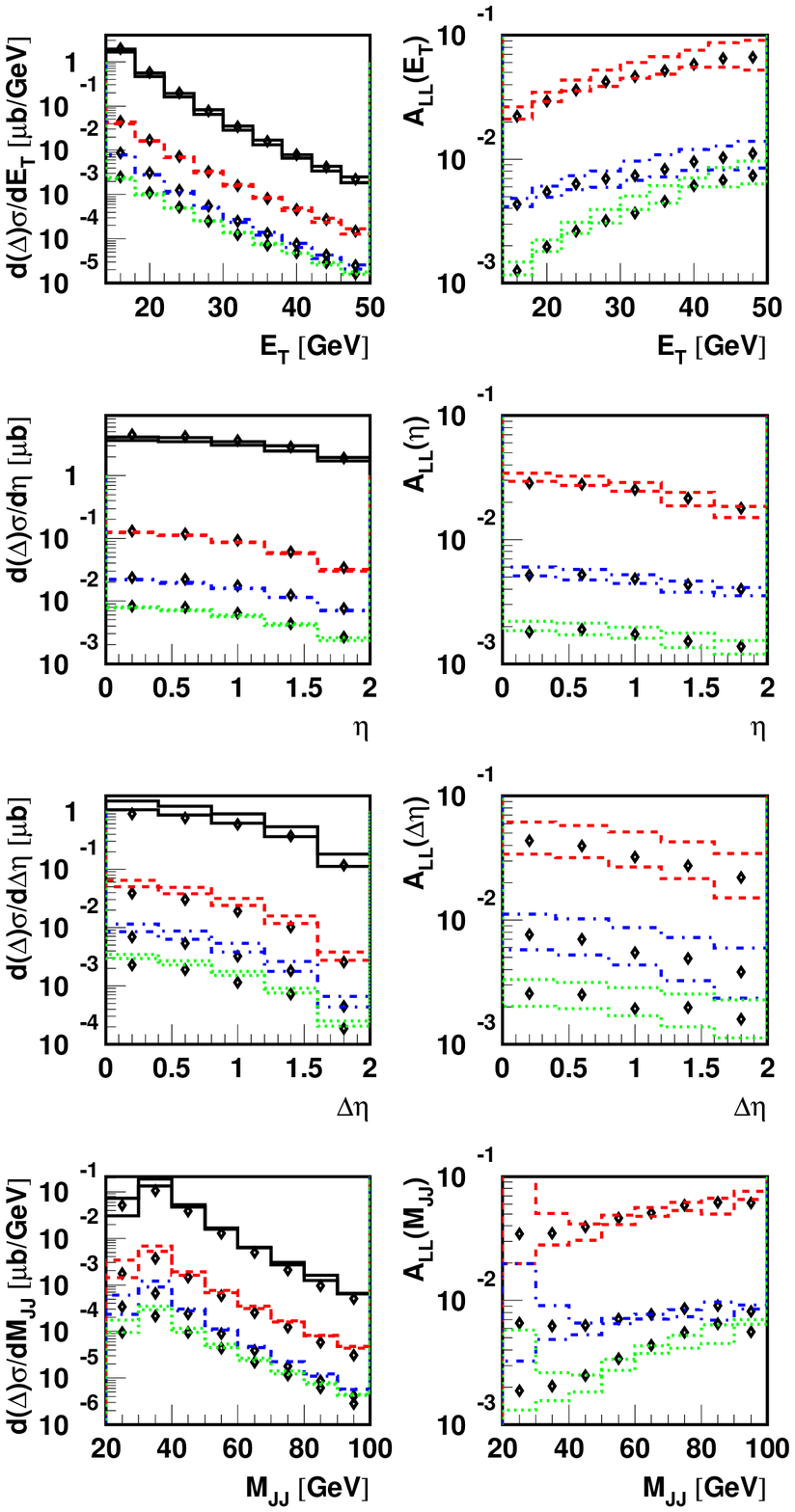,width=12cm}
\end{center}
\caption{\label{fig6}}
\end{figure}

\begin{figure}
\begin{center}
\epsfig{file=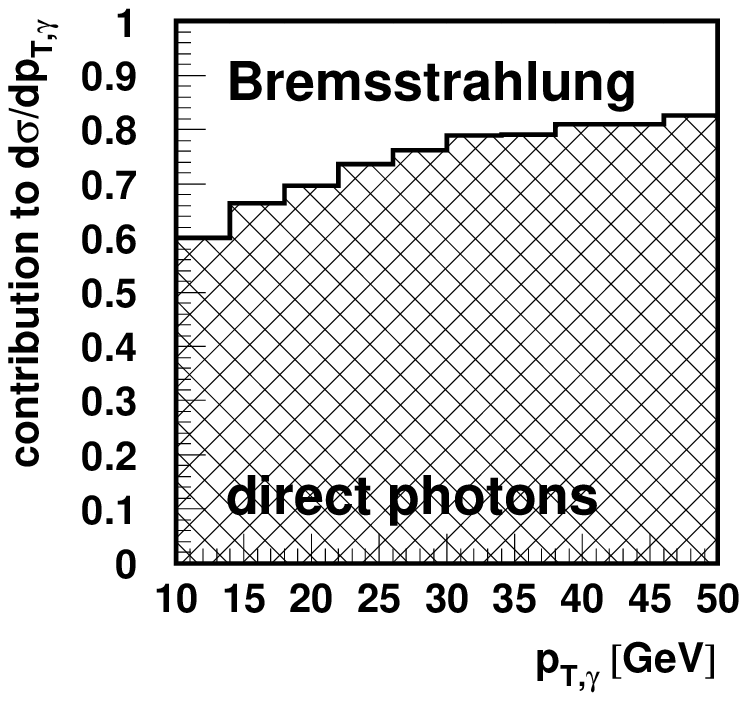,width=10cm}
\end{center}
\caption{\label{fig7}}
\end{figure}

\begin{figure}
\epsfig{file=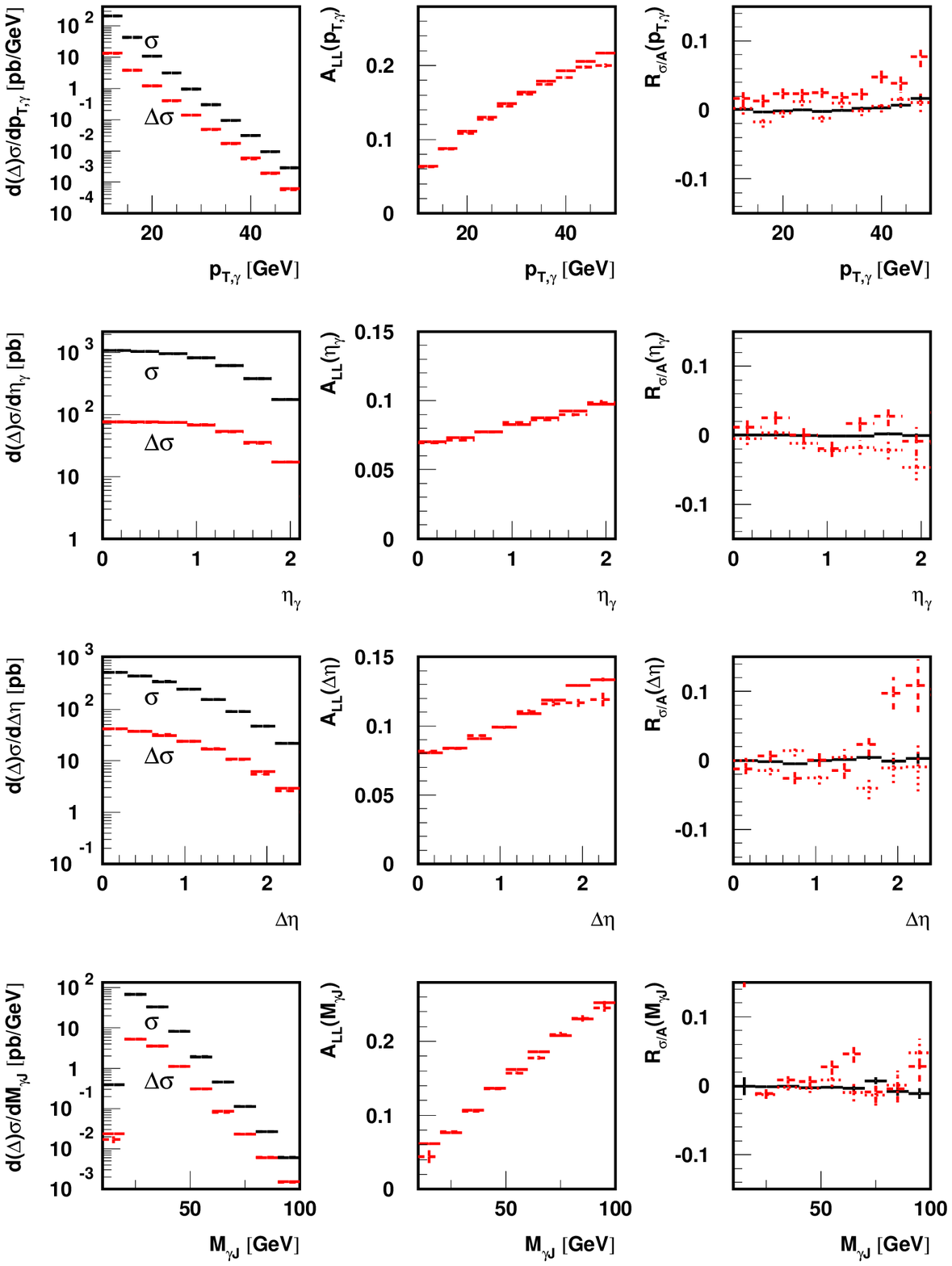,width=17cm}
\caption{\label{fig8}}
\end{figure}

\begin{figure}
\epsfig{file=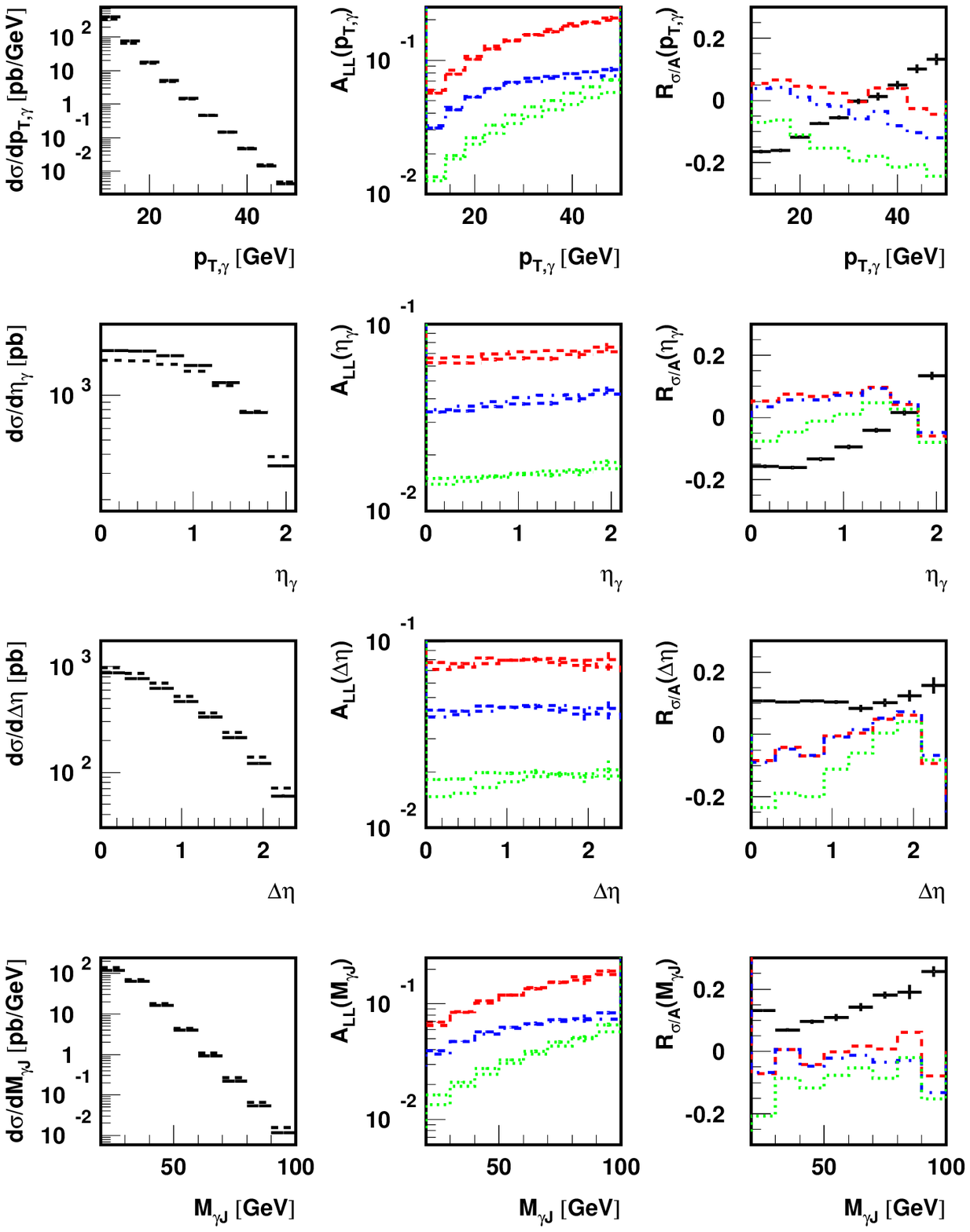,width=17cm}
\caption{\label{fig9}}
\end{figure}

\begin{figure}
\begin{center}
\epsfig{file=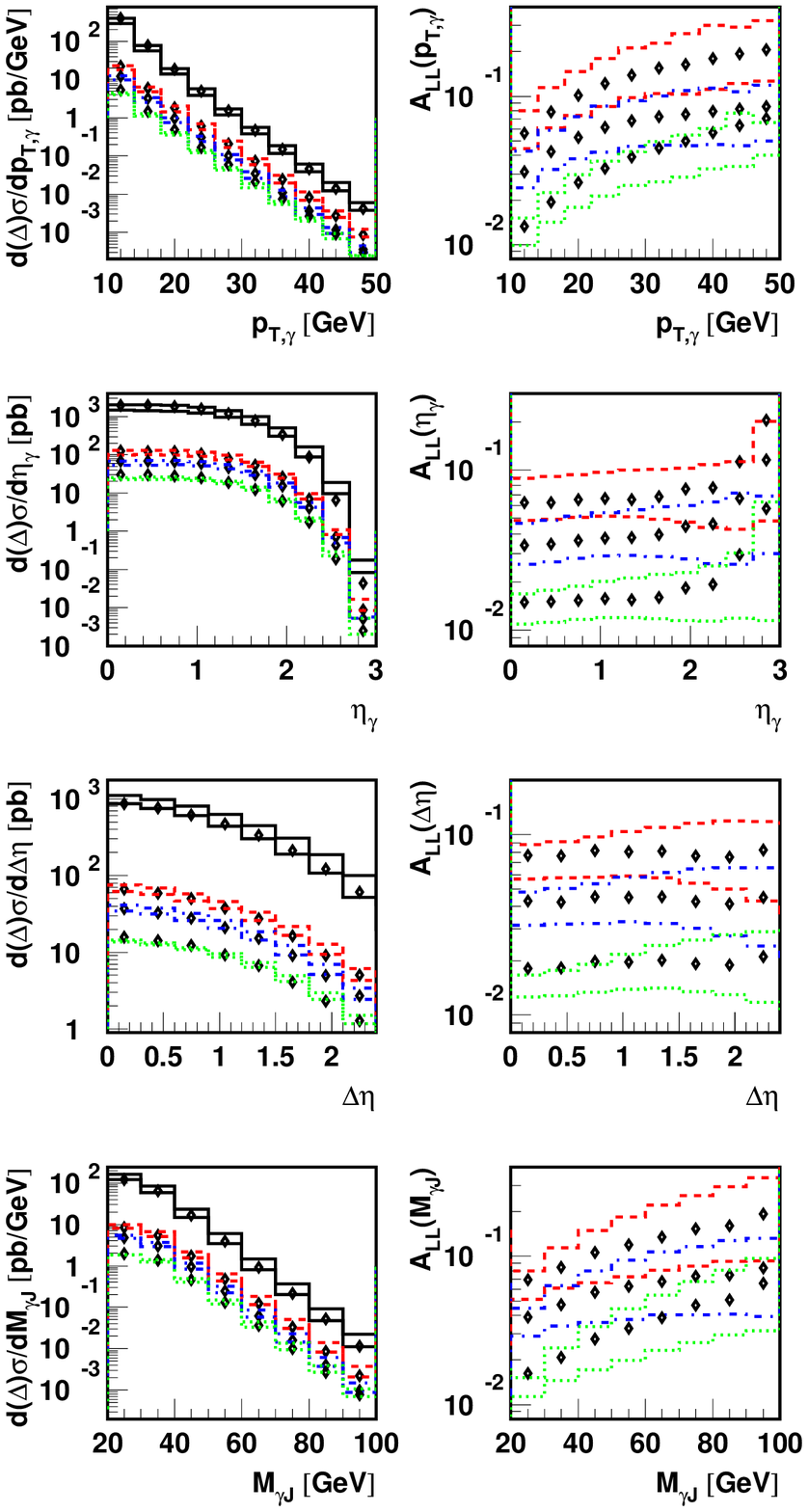,width=12cm}
\end{center}
\caption{\label{fig10}}
\end{figure}

\begin{itemize}
\item[\bf Fig. \ref{fig1}]
The ratio $K_{(\Delta)g}(x,\mu^2)$ defined in Eq.~(\ref{eq:kg}) for the 
unpolarized and polarized parton densities of our analysis in the relevant
$x$-range. The line style coding of this figure will be used
throughout the paper.
\item[\bf Fig. \ref{fig2}]
Comparison of jet production calculated with the MAW and {\sc Sphinx}: 
(un)polarized cross sections and asymmetries
calculated with GRSV~$g_{\rm max}$ and their relative deviations $R_{\sigma,A}$ defined in
Eq.~(\ref{eq:reldiff1}).  
Left and center columns: full lines represent the MAW results, dashed lines represent the
{\sc Sphinx} results. Right column: full lines represent $R_{\sigma}$, dashed (dotted) 
crosses represent 
$R_A$ calculated with (without) polarized treatment in the ISS of {\sc Sphinx}.
\item[\bf Fig. \ref{fig3}]
The relative deviations $R_{\sigma/A}$ defined in
Eq.~(\ref{eq:reldiff2}) for the Born results calculated with the MAW and the
parton generator. Line styles are chosen according to Fig.~\ref{fig1}.
The figure is based on $2\cdot  10^7$ generated events.
\item[\bf Fig. \ref{fig4}]
Comparison of jet production calculated with NLO QCD and the MAW with 
switched-on PS and standard PS parameters.
Left column: MAW (NLO~QCD) results are represented by full (dashed) lines.
Center column: the line styles are chosen according to Fig.~\ref{fig1} with one
line representing the NLO QCD and the other one the MAW result for
each parametrization. They can be identified from the third column, where
the line styles are again chosen according to Fig.~\ref{fig1},
i.e. full crosses represent $R_{\sigma}$ and all other lines represent $R_A$ for the
various parameterizations of polarized parton distributions. 
\item[\bf Fig. \ref{fig5}]
same as Fig.~\ref{fig4} but with tuned PS parameters which were chosen to
improve the agreement between both methods (see text).
\item[\bf Fig. \ref{fig6}]
Comparison of the optimized EG results (markers) with the bands of NLO QCD results which
are obtained, when the scales are varied by a factor of 2 or 1/2, resp.
Left column: full lines represent the band of possible unpolarized NLO QCD cross sections,
dashed and dotted lines represent the bands of polarized NLO QCD cross sections and 
follow the same pattern as in Fig.~\ref{fig1}. Right column: corresponding minimal and
maximal NLO QCD predictions for the asymmetries. Line Styles are again chosen 
according to Fig.~\ref{fig1}.
The large scale dependence for 20~GeV$<M_{JJ}<$30~GeV is due to the fact that
this bin receives no contribution to ${\cal O}(\alpha_s^2)$ so that all 
values are effectively only LO predictions.
\item[\bf Fig. \ref{fig7}]
Relative contribution of `direct' and `bremsstrahlung' photons to the total
production rate of isolated photons.
\item[\bf Fig. \ref{fig8}]
Same as Fig.~\ref{fig2} but for the production of `direct'
photons (without `bremsstrahlung' contribution).
\item[\bf Fig. \ref{fig9}]
Same as Fig.~\ref{fig4} but for the production of isolated photons (including the
`bremsstrahlung contribution').
A $K$-factor of $K=1.5$ is applied to all EG cross sections.
The figure is based on $3\cdot 10^8$ generated events.
\item[\bf Fig. \ref{fig10}]
Same as Fig.~\ref{fig6} but for the production of isolated photons.
A $K$-factor of $K=1.5$ is applied to all EG cross sections.
\end{itemize}
%
%
%
%
%
%
\newpage
\newpage
\newpage
\newpage
\newpage
\newpage
\newpage
\newpage
\newpage
\newpage
\end{document}